# Spin-orbit torque magnetization switching of a three-terminal perpendicular magnetic tunnel junction

Murat Cubukcu[1], Olivier Boulle[1], Marc Drouard[1], Kevin Garello[2], Can Onur Avci[2], Ioan Mihai Miron[1], Juergen Langer[3], Berthold Ocker[3], Pietro Gambardella[2] and Gilles Gaudin[1]

[1] *SPINTEC, UMR CEA/ CNRS, INAC, Grenoble, F-38054, France*

[2] *Department of Materials, ETH Zurich, Schafmattstrasse 30, CH-8093 Zürich, Switzerland*

[3] *Singulus Technologies, Hanauer Landstr, 103, 63796, Kahl am Main*, Germany

We report on the current-induced magnetization switching of a three-terminal perpendicular magnetic tunnel junction by spin-orbit torque and the read-out using the tunnelling magnetoresistance (TMR) effect. The device is composed of a perpendicular Ta/FeCoB/MgO/FeCoB stack on top of a Ta current line. The magnetization of the bottom FeCoB layer can be switched reproducibly by the injection of current pulses with density $5 \times 10^{11}$ A/m² in the Ta layer in the presence of an in-plane bias magnetic field, leading to the full-scale change of the TMR signal. Our work demonstrates the proof of concept of a perpendicular spin-orbit torque magnetic memory cell.

The microelectronics industry will face major challenges related to power dissipation and energy consumption in the next years. Both static and dynamic consumption (already dominated by the leakage power) of DRAM and SRAM cache memories will soon start to limit improvements in the performance of microprocessors [27]. A promising way to stop this trend is the integration of non-volatility in cache memories, which would minimize the static power consumption and pave the way towards normally-off/instant-on computing. The development of an electrically addressable non-volatile memory combining high speed and high endurance is essential to achieve these goals. Among emerging memory technologies, the Spin Transfer Torque Magnetic Random Access Memory (STT-MRAM) has been identified as one of the most promising candidates [1,28]. The STT-MRAM, however, suffers from serious reliability and endurance issues due to the rapid aging of the tunnel barrier induced by the high writing current densities as well as erroneous writing by the read current [23]. This is especially problematic in applications where speed is critical, such as cache memories, as larger writing current densities are required [4,19,23,24]. Scalability of the STT-MRAM beyond the 22 nm node is also under question, as the anticipated very low resistance area (R.A<1$\Omega.\mu m^2$) would lead to very challenging material optimizations to maintain a high tunneling magnetoresistance (TMR) [5]. These problems complicate the realization of STT-MRAM, particularly for applications where scalability and speed are keys.

Recently, an alternative way to switch the magnetization of a thin ferromagnetic layer using an electric current has been demonstrated based on the concept of the spin-orbit torque (SOT) [6,7,8]. In this scheme, the magnetic free layer is in contact with a nonmagnetic heavy metal stripe, typically Pt or Ta. When injecting a current in the nonmagnetic layer, the spin-orbit coupling leads to a perpendicular spin current which is transferred to the magnetization creating a spin torque and inducing magnetization reversal [6,8]. This spin current may arise from the spin Hall effect and Rashba-like interface effects, their relative contributions being currently under debate [6,8,20-22].

A concept for a three-terminal magnetic memory device based on this discovery has been proposed, namely the SOT-MRAM, where the magnetic bit is written by a current pulse injected through the bottom metallic electrode and a MTJ is used to read the state of the magnetic bit [9]. The key advantage of the SOT-MRAM is that the write and read operations are decoupled due to the different current paths, which naturally solves the problems related to endurance and read disturbance of the STT-MRAM. Moreover, the TMR can be tuned independently of writing constraints, relaxing the highly demanding material requirement of combining very low R.A and high TMR.

Recently, three-terminal devices with in-plane magnetized MTJ patterned onto Ta [8], Ir-doped Cu [10], and W [11] electrodes have been demonstrated. However, materials with high perpendicular magnetic anisotropy are more suitable for memory applications due to their ability to retain a stable magnetization state over long periods of time. In addition, in the presence of an in-plane bias magnetic field, their magnetization can be reversed by extremely short current pulses (<400 ps) in this geometry, since the SOT is perpendicular to the quiescent magnetization direction, leading to writing energies expected below 100 fJ at 100 nm lateral dimensions [12]. In this letter, we demonstrate the proof of concept of a perpendicular SOT-MRAM cell composed of a FeCoB/MgO/FeCoB MTJ deposited on top of a 20 nm thick Ta current line. The basic write and read operations, i.e., the magnetization reversal by current injection in the Ta electrode and its detection using the TMR signal, are demonstrated. Our results open the way to develop three-terminal non-volatile magnetic memories with perpendicular magnetization, which can combine fast and low energy writing with high endurance.

The magnetic tunnel junction is composed of a Ta(20nm)/Fe$_{60}$Co$_{20}$B$_{20}$(1nm)/MgO/Fe$_{60}$Co$_{20}$B$_{20}$(1.5nm)/Ta(5nm)/Ru(7nm) stack that was deposited on a high resistivity Si substrate by magnetron sputtering using a Singulus Timaris© deposition machine and annealed at 240 °C for 1h. The three-terminal devices are composed of circular MTJ dots on top of the Ta track (Fig. 1 (a)). They were fabricated as follows. A Ru(5nm)/Ta(150nm) layer was first deposited on the MTJ stack using a magnetron sputtering tool. A Ta nanopillar was then patterned by e-beam lithography and anisotropic reactive ion etching (RIE). The MTJ dot was then defined by ion beam etching, the Ta pillar being used as a mask. The bottom Ta electrode was then fabricated by patterning a resist mask by e-beam lithography followed by an IBE step. Finally, a planarization resist was spin coated and etched by RIE until the top of the Ta nanopillar emerged. The top Cr/Al electrodes were then successively defined by optical lithography and evaporation. Figure 1 (b) shows a scanning electron microscopy image of a circular MTJ with 1µm diameter on top of a 1.3 µm wide Ta electrode patterned in a Hall cross shape. The Hall cross allows for the electrical detection of the orientation of the bottom layer magnetization using the anomalous Hall effect (AHE). The diameters of the dots studied here are 200 nm, 500 nm, and 1 µm. The resistance of the MTJ is measured by injecting a small DC current ($I$=2 µA). To prevent current injection in the MTJ when injecting a current in the bottom Ta layer, a 100 kΩ resistance was connected in series with the MTJ close to the sample. All measurements are performed at room temperature.

Figure 2 (a) shows the TMR of a 1 µm diameter MTJ as a function of magnetic field $H$ applied perpendicularly to the sample plane. The abrupt TMR change during the field sweep shows that both magnetic layers are perpendicularly magnetized. A simultaneous measurement of the AHE shows that the top electrode switches first when sweeping the magnetic field. From the shift of the minor hysteresis cycle of the top FeCoB layer (Fig. 2 (a), red curve), we estimate that the bottom layer exerts a dipolar field of about 75 Oe on the top layer. This sample exhibits a TMR signal of 55 % and an R.A product of 1.15 kΩ.µm², but TMR up to 90 % were measured in other samples patterned from the same stack.

Figure 2 (b) demonstrates magnetization switching by an in-plane current of a three terminal device and its detection by TMR. The TMR signal is measured after the injection of a 50 ns current pulse in the Ta track. A constant external magnetic field of $H_l =$-0.4 kOe is applied along the current direction to determine the switching polarity [6]. This field has a small perpendicular component (≈5°) which favors a well-defined orientation of the top FeCoB layer and prevents the formation of magnetic domains. Starting from a parallel configuration, the injection of a large enough positive current pulse leads to an increase of the TMR signal, revealing the reversal of the magnetization of the bottom FeCoB layer. From the antiparallel configuration the magnetization can be switched back to a parallel configuration by injecting a negative current pulse. The bottom FeCoB magnetization can thus be switched hysteretically by current pulse injection in the bottom Ta electrode in the presence of $H_l$ and its orientation detected by the TMR signal.

As shown on Figure 3 (a, c), the resistance of the device can be switched back and forth and reproducibly between the two TMR states, by injecting successive current pulses in the Ta current line with opposite polarities ($I_p$=20 mA, 50 ns-long pulses, $H_l$=-0.4 kOe). As expected, a simultaneous oscillation of the anomalous Hall signal in the Hall cross is observed, which confirms that switching is due to the reversal of the bottom FeCoB layer (Fig. 3 (b)).

A similar switching behavior is observed for dots with smaller diameter. The critical switching current is estimated to be $I_c$=9 mA ($j_c$≈6x10$^{11}$ A/m²) for 500 nm dots on top a 770 nm wide Ta electrode and $I_c$=6 mA ($j_c$≈6x10$^{11}$ A/m²) for the 200 nm MTJ on top a 470 nm wide Ta electrode at $H_1=-0.4$ kOe. There is thus little dependence of the critical current density on the dot size in this range and the critical current scales with the cross section of the bottom electrode.

More insight is obtained when sweeping the magnetic field at constant pulse amplitude. Figure 4 shows the AHE resistance (Fig. 4 (a)) and the TMR signal (Fig. 4 (b)) as a function of $H_1$ after the successive injection of a positive (black square) or negative (red dots) current pulse of amplitude $I=20$ mA ($j \approx 7.5 \times 10^{11}$ A/m²) for a dot of 1 µm diameter. At low $|H_1|<0.1$ kOe, no magnetization switching is observed, which shows that neither the external field nor the in-plane component of the stray field from the top layer is sufficient to trigger the magnetization reversal. For 0.1 kOe $< H_1 <$ 0.7 kOe, current induced magnetization switching occurs: the injection of positive (resp. negative) current pulse leads to the reversal of the bottom FeCoB layer magnetization along the up (resp. down) direction (AHE signal, Fig.4 (a)). Since the magnetization of the top electrode points upward in this field range due to the perpendicular component of $H_1$, the switching of the bottom FeCoB layer leads to different TMR signals for positive and negative pulses (Fig. 4(b)). For negative magnetic field, switching of the bottom FeCoB layer is also observed but with reversed polarity of current pulses.

These experiments can be used to draw a current-field switching diagram constructed using the TMR signal for a 1 µm diameter MTJ dot (Fig. 5(a)). Different behaviors are observed when injecting current pulses depending on the value of $j$. At low current density $j<5\times10^{11}$ A/m² (region I), no hysteretic switching is observed, but current injection leads to a reduction of the coercive field of the bottom layer (black open-triangles) with increasing pulse amplitude. This can be understood by the combination of a weak current-induced effect and the residual vertical component of the external field $H_z$, which assists magnetization reversal towards the equilibrium direction, parallel to $H_z$ [6]. At higher current density ($j>5\times10^{11}$ A/m²) (region III), current-induced hysteretic switching occurs. The associated $j$-$H_l$ region (III, green) is delimited at low $H_l$ by the minimum magnetic field required to trigger the switching and at high $H_l$ by the coercive field of the bottom layer due to the perpendicular component of $H_l$. In between these two regions, no antiparallel state is observed (region II, red). In this case, the reversal of the bottom FeCoB layer occurs at the coercive field of the top layer and is promoted by the stray field of the reversed magnetization of the top layer. Note that this switching diagram is similar to the one measured for a single layer [6], except for region II where no change in the TMR signal is observed.

We also studied the dependence of the critical current $I_c$ on the pulse length $\tau_p$ for 30 ns$<\tau_p<$500 ns (Fig. 5(b), $H_l$=-0.4 kOe). We observe that $I_c$ scales linearly with $1/\tau_p$ for $\tau_p<$100 ns. A similar scaling has been reported in Pt/Co/AlO$_x$ trilayer [6] down to 400 ps [12]. This result is in contradiction with macrospin simulations, which predict a weak dependence of $I_c$ on $\tau_p$ [12, 13]. Such a scaling is more consistent with a reversal process governed by domain nucleation followed by domain wall (DW) motion, which is expected given the large diameter of the pillar [14]. Assuming that the DW velocity scales linearly with the current density, the critical current required for the DW to traverse the whole pillar is expected to scale as $1/\tau_p$, as observed experimentally.

The critical current density for which the magnetization can be reversed back and forth by current pulses with opposite polarities ($\approx 5 \times 10^{11}$ A/m² at 50 ns) is higher compared to $j_c$ reported in similar experiments in perpendicularly magnetized Ta/CoFe/MgO [8] ($j_c$=8x10$^{10}$ A/m$^2$) and Ta/CoFe/MgO [15] ($j_c$=1x10$^{10}$ A/m$^2$) multilayers. This may be accounted for by several features of our experimental scheme. First, we inject short current pulses (50 ns) and use a Si substrate, whereas previous experiments used quasi DC current and Si/SiO$_2$ substrates, which result in a smaller critical current density due to thermally assisted magnetization reversal. Second, part of the current flows into the arms of the Hall cross which leads to a local decrease of the current density below the MTJ dots, resulting in a lower torque [16]. Third, the dipolar field exerted by the top layer and the perpendicular component of the external magnetic field prevent the switching to an antiparallel configuration and thus increase the critical current density for which bipolar switching is observed.

A key parameter for the density of the memory is also the writing current that will define the size of the addressing transistor. For $j_c = 5 \times 10^{11}$ A/m², the large thickness and width of the bottom Ta stripe lead to relatively high critical switching current in our experiments. Assuming that the Slonczewski-like part of the SOT arises mainly from the spin Hall effect [22], the spin current generated by current injection should saturate for Ta layers thicker than the spin diffusion length [26]. This was estimated between to be about 1.8 nm in Ta at room temperature [25]. In addition, the critical current density should remain approximately constant as a function of dot size. Thus, assuming a 3 nm thick 50 nm wide Ta track, a critical current of about 25 µA is expected which is similar to the best results for perpendicular STT-MRAM [17,18,19]. Finally, we note that the longitudinal magnetic bias field can be generated by a set of permanent magnets [9].

In summary, we have fabricated a three-terminal elementary memory cell of a perpendicular SOT-MRAM and demonstrated its basic read and write operations. The SOT-MRAM cell is composed of a perpendicular FeCoB/MgO/FeCoB MTJ with high TMR on top of a Ta track. We have shown that the magnetic bit represented by the relative alignment of the magnetization of the top and bottom FeCoB layers can be reproducibly written by the injection of current pulses in the Ta bottom electrode ($5 \times 10^{11}$ A/m², 50 ns-long pulses) and read by measuring the large TMR signal. Our work demonstrates the proof of concept of a perpendicular SOT-MRAM memory cell.

This work was supported by the European Commission under the Seventh Framework Program (Grant Agreement 318144, spOt project). The devices were fabricated at the Plateforme de Technologie Amont (PTA) in Grenoble.

**References**

[1] J. Hutchby, and M. Garner, Assessment of the Potential and Maturity of Selected Emerging Research Memory Technologies Workshop and ERD/ERM Working Group Meeting, 6–7 April 2010, p. 1 (ITRS Edition) (2010)

[2] W. Zhao, E. Belhaire, C. Chappert, and P. Mazoyer, ACM Trans. Embed. Comput. Syst. **9** 14:1–14:16 (2009)

[3] G. Prenat, M. El Baraji, W. Guo, R. Sousa, L. Buda-Prejbeanu, B. Dieny, V. Javerliac, J.-P. Nozieres, W. Zhao, and E. Belhaire, In 14th IEEE International Conference on Electronics, Circuits and Systems, ICECS, **190** (2007)

[4] M. Marins de Castro, R. C. Sousa, S. Bandiera, C. Ducruet, A. Chavent, S. Auffret, C. Papusoi, I.L. Prejbeanu, C. Portemont, L. Vila, U. Ebels, B. Rodmacq, and B. Dieny, J. Appl. Phys. Lett. **111**, 07C912 (2012)

[5] K.C. Chun, H. Zhao, J. Harms, T.-H. Kim, J.-P. Wang, and C. Kim, IEEE Journal of Solid-State Circuits **48**, 598 (2013)

[6] I. M. Miron, K. Garello, G. Gaudin, P.-J. Zermatten, M.V. Costache, S. Auffret, S. Bandiera, B. Rodmacq, A. Schuhl, P. Gambardella , Nature **476**, 189 (2011)

[7] C. O. Avci, K. Garello, I.-M. Miron, G. Gaudin, S. Auffret, O. Boulle, Appl. Phys. Lett., **100**, 212404 (2011)

[8] L. Liu, C. Pai, Y. Li, H. M. Tseng, D.C. Ralph, and R.A. Buhrman, Science **336**, 55 (2012)

[9] G. Gaudin, I. M. Miron, P. Gambardella, A. Schuhl, Magnetic memory element, Patent, US Patent application, 12/899,072, 12/899,091,12/959980, (2010)

[10] M. Yamanouchi, L. Chen, J. Kim, M. Hayashi, H. Sato, S. Fukami, S. Ikeda, F. Matsukura, H. Ohno, Appl. Phys. Lett. **102**, 212408 (2013)

[11] C. Pai, L. Liu, Y. Li, H. M. Tseng, D.C. Ralph, and R.A. Buhrman, Appl. Phys. Lett. **101**, 122404 (2012)

[12] K. Garello, C.O. Avci, I.M. Miron, O. Boulle, S. Auffret, P. Gambardella and G. Gaudin, arXiv:1310.5586 (2013)

[13] K.-S. Lee, S.-W. Lee, B.-C. Min and, K.-J. Lee, Appl. Phys. Lett., **102**, 112410 (2013)

[14] G. Finocchio, M. Carpentieri, E. Martinez, B. Azzerboni, Appl. Phys. Lett. **102**, 212410 (2013)

[15] S. Emori, U. Bauer, S-M. Ahn, E. Martinez, and G. S. D. Beach. Nat. Mater., **12**, 611 (2013)

[16] K. Garello, I. M. Miron, C. O. Avci, F. Freimuth, Y. Mokrousov, S. Blügel, S. Auffret, O. Boulle, G. Gaudin, P.Gambardella, Nat. Nano. , **8**, 587 (2013)

[17] K. Yamane, Y. Higo, H. Uchida, Y. Nanba, S. Sasaki, H. Ohmori, K. Bessho and M. Hosomi, IEEE Transactions on Magnetics **49** (7), 4335–4338 (2013)

[18] G. Jan, Y-J. Wang, T. Moriyama, Y-J Lee, M. Lin, T. Zhong, R-Y. Tong, T. Torng, P-K. Wang, Appl. Phys. Exp., **5**, 93008 (2012)

[19] M. Gajek, J. J. Nowak, J. Z. Sun, P. L. Trouilloud, E. J. O'Sullivan, D. W. Abraham, M. C. Gaidis, G. Hu, S. Brown, Y. Zhu, R. P. Robertazzi, W. J. Gallagher and D. C. Worledge, Appl. Phys. Lett, **100**, 132408 (2012)

[20] P.M. Haney, H.-W. Lee, K.-J. Lee, A. Manchon and M.D. Stiles, Phys. Rev. B **87**, 174411 (2013)

[21] F. Freimuth, S. Blugel, Y. Mokrousov, arXiv:1305.4873 (2013)

[22] P.M. Haney, H.-W. Lee, K.-J. Lee, A. Manchon and M.D Stiles, arXiv:1309.1356 (2013)

[23] W.S. Zhao, Y. Zhang, T. Devolder, J.O. Klein, D. Ravelosona, C. Chappert, P. Mazoyer, Microelectronics Reliability, **52**, 1848-1852 (2012)

[24] G. Panagopoulos, C. Augustine, K. Roy, In: Proc in device research conference (DRC) p. 125–6 (2011)

[25] C. Hahn, G. de Loubens, O. Klein, and M. Viret, V. V. Naletov, J. Ben Youssef, Phys. Rev. B, **87,** 174417 (2013)

[26] L. Liu, T. Moriyama, D. C. Ralph, and R. A. Buhrman, Phys. Rev. Lett., **106**, 036601 (2011)

[27] N. S. Kim, T. Austin, D. Baauw, T. Mudge, IEEE Computer Society, pp. 68-74 (2003)

[28] M. H. Kryder and C. Soo Kim, IEEE Transactions on magnetics, **45**, 10 (2009)

**Figure Captions:**

Fig.1 (a) Illustration of a three-terminal MTJ with a Hall cross geometry. The black (blue) 'up' and red (yellow) 'down' arrows indicate the equilibrium magnetization states of the bottom (top) FeCoB layers. (b) Scanning electron microscopy of a 1µm dot diameter patterned MTJ on top a 1.3µm wide Ta electrode with a schematic representation of the electrical measurement setup. $V_{Hall}^{+}$, $V_{Hall}^{-}$ and $V_{MTJ}$ represent the two terminals for the Hall voltage and MTJ voltage measurements respectively.

Fig. 2 (a) TMR as a function of the perpendicular magnetic field for 1µm dot diameter. The black and red curves show respectively a major and a minor loop. (b) TMR as a function of current pulse amplitude $I_p$ injected in the Ta electrode using 50 ns long pulses under an in-plane magnetic field $H_l$=-0.4 kOe. The arrows show the sweep direction of $I_p$.

Fig. 3 (a) Schematic of the pulse sequence. (b) The AHE resistance (proportional to the $M_z$ component of the bottom FeCoB layer) and (c) TMR measured after the injection of positive (black squares) and negative (red circles) current pulses of amplitude $I_p$=20 mA and 50 ns long under $H_l$=-0.4 kOe.

Fig. 4 (a) The AHE resistance and (b) TMR as a function of $H_l$ measured after the injection of positive (black squares) and negative current pulses (red circles) of amplitude $I_p$=20 mA and a length of 50 ns. The dot diameter is 1µm. $H_l$ is swept from negative to positive magnetic fields.

Fig. 5 (a) $I_P$-$H_l$ switching diagram constructed from the measurement of the TMR signal for a 1 µm diameter MTJ (50 ns long pulses). Region I: current assisted magnetization reversal. The open black up-triangles and the dotted line indicate the coercive field of the bottom and top FeCoB layers respectively. Region II: low TMR signal (parallel alignment). Region III: Current-induced magnetization switching. The red squares and the blue circles indicate the maximum field and minimum field at which current induced magnetization switching of the bottom layer is observed. $P_\uparrow$ and $P_\downarrow$ indicate the parallel state where both FeCoB magnetizations are oriented upward and downward respectively. (b) Dependence of the critical current $j_\text{c}$ on the inverse pulse width ($1/\tau_\text{p}$). The black line is a guide for the eyes.

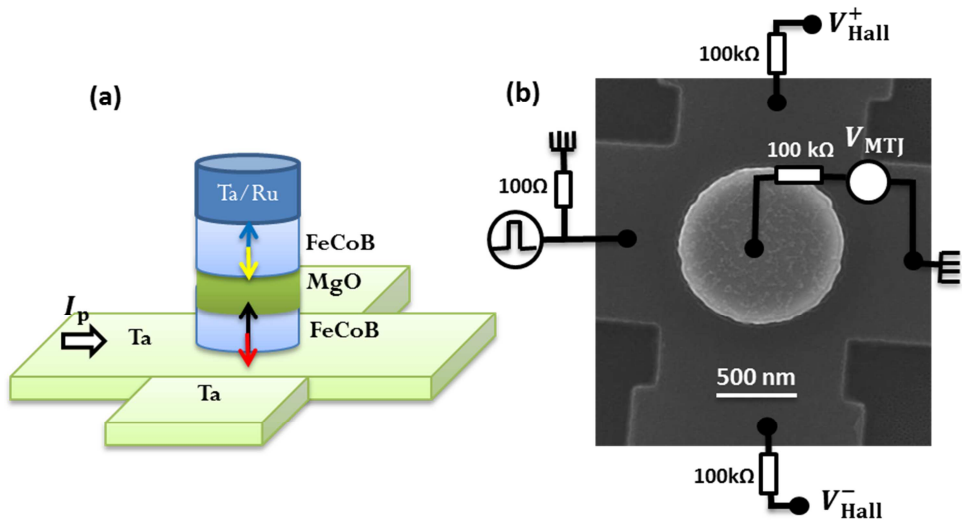

Fig.1

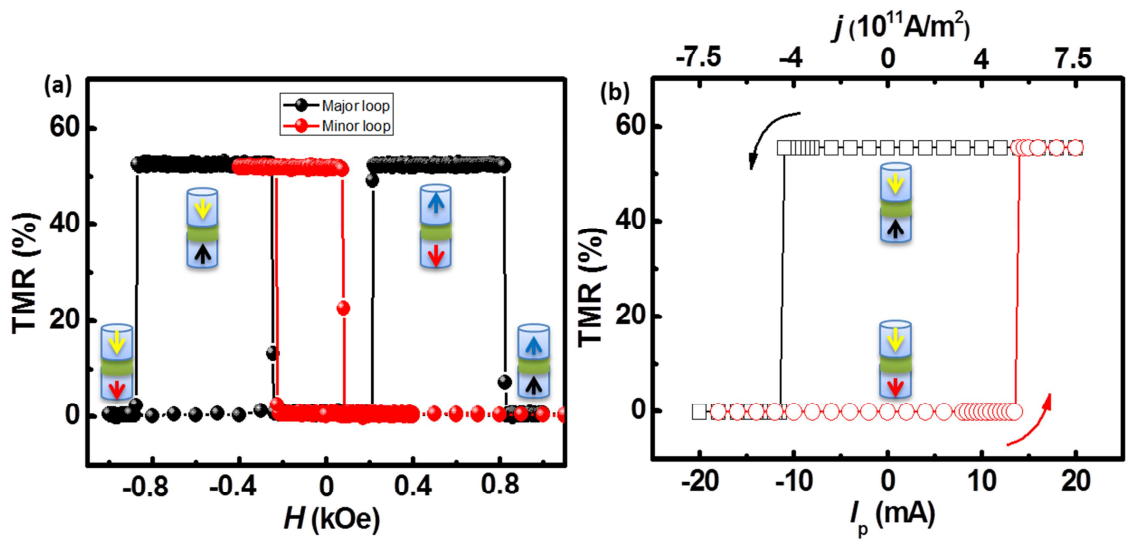

Fig.2

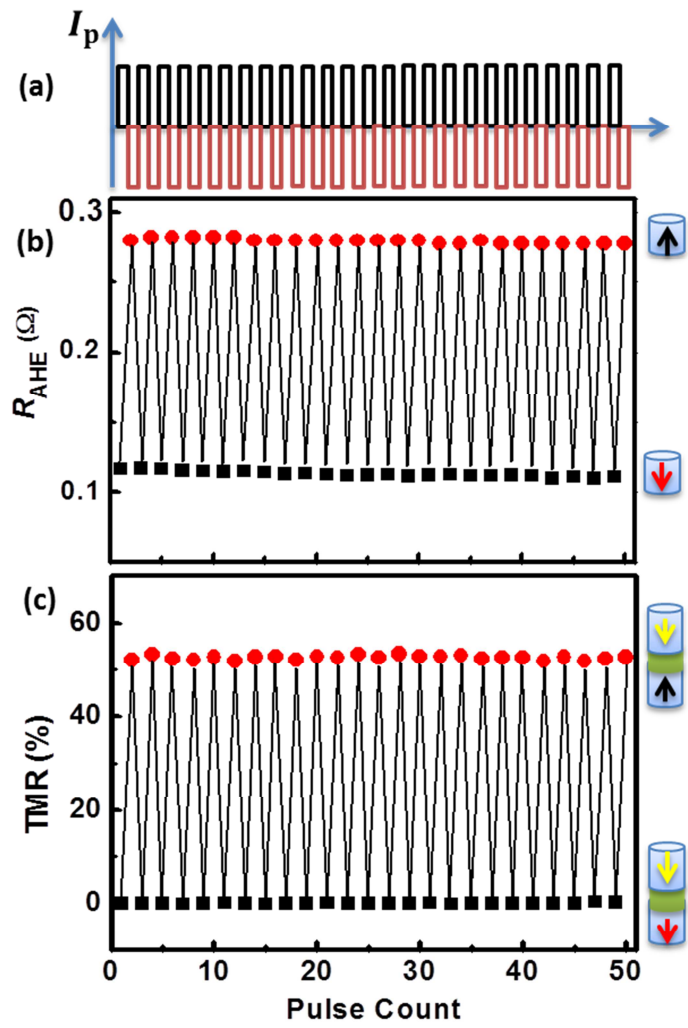

Fig.3

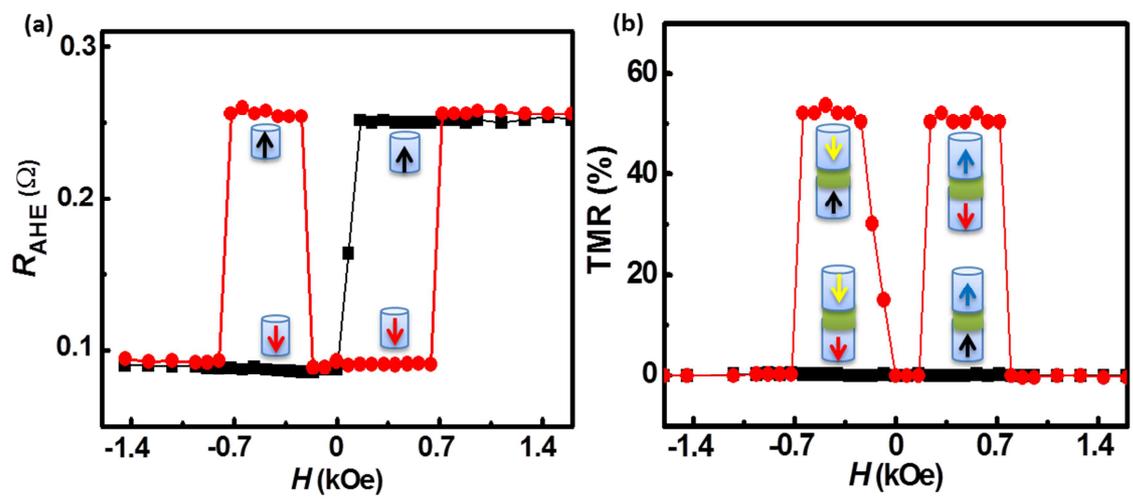

Fig.4

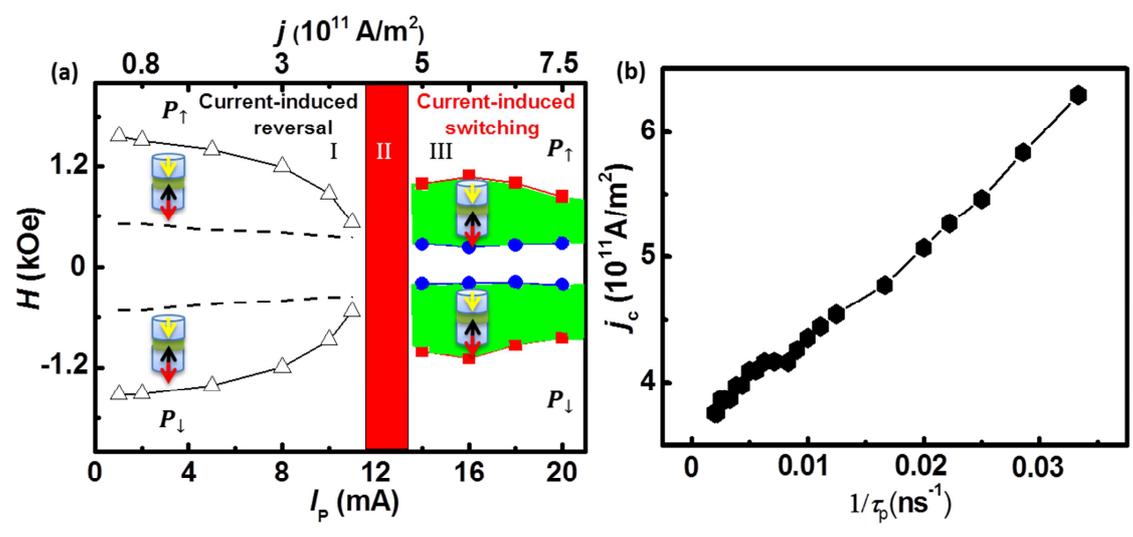

Fig.5

17